\documentclass[twocolumn,english,prb, aps, showpacs]{revtex4}

\usepackage[latin1]{inputenc}
\usepackage{amsmath}
\usepackage{graphicx}
\usepackage{amssymb}

\makeatletter
\usepackage{bm}
\newcommand{\nn}{\nonumber}
\newcommand{\al}{\alpha}

\newcommand{\la}{\langle}
\newcommand{\ra}{\rangle}

\newcommand{\up}{\uparrow}
\newcommand{\down}{\downarrow}
\newcommand{\E}{\mathcal{E}}

\usepackage{babel}
\makeatother
\begin{document}

\title{Spin Relaxation in a Quantum Dot due to Nyquist Noise}

\author{Florian Marquardt$^{1}$}

\email{Florian.Marquardt@yale.edu}

\author{Veniamin A. Abalmassov$^{2}$}

\email{V.Abalmassov@isp.nsc.ru}

\affiliation{$^{1}$Department of Physics, Yale University, New Haven CT, 06520,
USA}

\affiliation{$^{2}$Institute of Semiconductor Physics SB RAS and Novosibirsk
State University, 630090 Novosibirsk, Russia}

\date{\today{}}

\begin{abstract}
We calculate electron and nuclear spin relaxation rates in a quantum
dot due to the combined action of Nyquist noise and electron-nuclei
hyperfine or spin-orbit interactions. The relaxation rate is linear
in the resistance of the gate circuit and, in the case of spin-orbit
interaction, it depends essentially on the orientations of both the
static magnetic field and the fluctuating electric field, as well
as on the ratio between Rashba and Dresselhaus interaction constants.
We provide numerical estimates of the relaxation rate for typical
system parameters, compare our results with other, previously discussed
mechanisms, and show that the Nyquist mechanism can have an appreciable
effect for experimentally relevant systems.
\end{abstract}


\maketitle

\section{Introduction}

The dynamics of electron and nuclear spins in quantum dots has become
a focus of research recently, with part of the motivation stemming
from the prospect of potential applications in the context of spintronics
\cite{Wolf01} or quantum computation \cite{lossdavid,Nielsen00}.
Of particular importance are relaxation processes leading to decoherence,
which have received much theoretical \cite{KhN01,Erlingsson02,Golovach03,AM,Kane01,Wellard02}
and experimental \cite{FujisawaPRB01,FujisawaNat02,HansonPRL03,Huettel03,KouwenhovenNature04,KrutvarNature}
attention within the last few years. 

Relaxation between Zeeman-split spin levels involves energy dissipation,
which, in quantum dots with their discrete orbital spectrum, requires
coupling to external degrees of freedom (a bath). In most theoretical
works, phonons have played the role of such a bath \cite{KhN01,Erlingsson02,Golovach03}.
Here, we will consider the Nyquist fluctuations of electric fields
produced by nearby gate electrodes as an alternative source of dissipation.
Decoherence due to Nyquist noise has so far been considered extensively
only for charge-based quantum computation proposals \cite{averin,makhlin,Dykman}.
The potential contribution to the spin relaxation rate of magnetic
field fluctuations (due to Nyquist current noise) has been found to
be very small\cite{KhN01}. More recently, decoherence of phosphorous
electron and nuclear spins in silicon due to Nyquist-noise induced
fluctuations of the hyperfine constant has been analyzed in Refs.
\onlinecite{Kane01,Wellard02}. 

\begin{figure}
\includegraphics[%
  width=0.90\columnwidth]{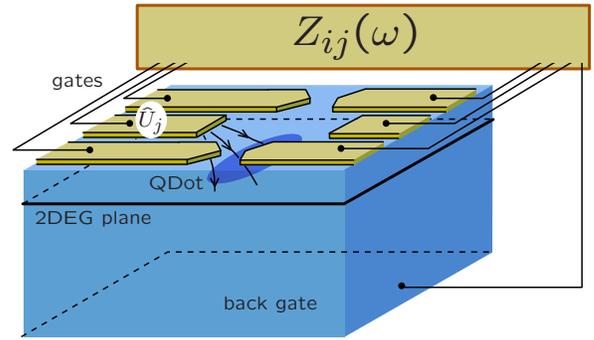}

\caption{\label{cap:Typical-gated-lateral}(Color online) Typical gated lateral
quantum dot structure considered in the text, with gates connected
to an electrical circuit of impedance matrix $Z$. The fluctuating
electric field produced by one particular gate is shown schematically.}
\end{figure}

In the present work, we will calculate the electron spin relaxation
rate in a single lateral quantum dot (Fig. \ref{cap:Typical-gated-lateral}),
due to Nyquist fluctuations of a gate voltage, combined either with
electron-nuclei hyperfine interaction or spin-orbit interactions.
In the case of hyperfine coupling, this also contributes to nuclear
spin relaxation within the quantum dot. In the case of spin-orbit
coupling, the spin relaxation rate displays a striking dependence
on the directions of both the fluctuating electric field as well as
the static magnetic field. The spin-orbit Nyquist mechanism considered
in this work can become as efficient as coupling to piezoelectric
phonons (the most important other mechanism) in realistic experimental
setups. The rate grows linearly in resistance of the gate circuit. 

\begin{figure}
\includegraphics[%
  width=0.80\columnwidth]{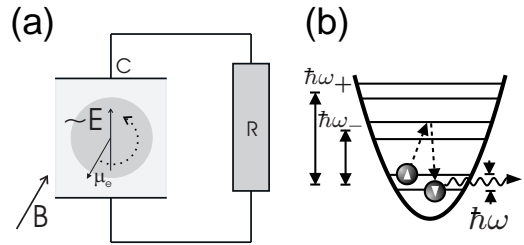}

\caption{(a) Equivalent circuit scheme used in our model; (b) Energy diagram,
showing the second-order relaxation process releasing the Zeeman energy
$\hbar\omega$ into the electromagnetic environment.}

\label{fig:model}
\end{figure}

In the following, we will first introduce our model Hamiltonian (Sec.
\ref{sec:The-model}), then describe the general calculation of spin
flip rates in second order Golden Rule (Sec. \ref{sec:Spin-flip-rate}),
and apply this scheme to both the hyperfine (Sec. \ref{sec:Hyperfine-interaction})
and spin-orbit (Sec. \ref{sec:Spin-orbit-interaction}) interactions.
We will comment on the relation to recent spin relaxation measurements\cite{Huettel03,KouwenhovenNature04}.
Finally, we will explain how the results may be applied to arbitrary
gate geometries and impedances (Sec. \ref{sec:Arbitrary-gate-geometries})
and comment on deviations due to other dot potentials or spatially
inhomogeneous electric field fluctuations (Sec. \ref{sec:Other-dot-shapes}).

\section{The model}

\label{sec:The-model}We start by considering a single electron in
a quantum dot formed in a 2DEG by a circularly symmetric parabolic
lateral confining potential, subject to a homogeneous static magnetic
field ${\mathbf{B}}$: \begin{eqnarray}
\hat{H}_{0}=\frac{\hat{\bm\pi}^{2}}{2m}+\frac{m\Omega_{0}^{2}}{2}{\bm\rho}^{2}+V(z)+\text{g}\mu_{B}{\hat{\mathbf{S}}}\cdot{\mathbf{B}},\label{hamilt0}\end{eqnarray}
 where ${\hat{\bm\pi}}={\mathbf{\hat{p}}}-(q/c){\mathbf{A}}$ is the
kinetic electron momentum, $q=-|e|$ the electron charge, ${\bm\rho}$
the in-plane position vector, $m$ and g the effective electron mass
and g-factor respectively, $\mu_{B}>0$ the Bohr magneton, $V(z)$
the transverse confining potential, and $\Omega_{0}$ the lateral
frequency.

The solutions of the in-plane orbital part are Fock-Darwin states
$|n_{+}n_{-}\ra$ with energies $n_{+}\hbar\omega_{+}+n_{-}\hbar\omega_{-}$,
where $\omega_{\pm}=\sqrt{\Omega_{0}^{2}+(\omega_{c}/2)^{2}}\pm\omega_{c}/2\equiv\Omega\pm\omega_{c}/2$
and the cyclotron frequency is $\omega_{c}=|q|B_{z}/(mc)$. We will
need the orbital ground state $\phi_{00}$ and the wave functions
of the first excited doublet, $\phi_{00}=(\pi l)^{-1/2}\exp\left[-\rho^{2}/(2l^{2})\right]$
and $\phi_{10}=\phi_{01}^{*}=(\rho e^{i\varphi}/l)\phi_{00}$, with
$x+iy=\rho e^{i\varphi}$ and $l=\sqrt{\hbar/(m\Omega)}$.

We now consider spin relaxation due to the Nyquist fluctuations of
an electric field acting onto the electron, being produced by a nearby
gate (see Figs.~\ref{cap:Typical-gated-lateral} and \ref{fig:model}).
We will at first assume the direction of the fluctuating field to
be fixed (Fig.~\ref{fig:model}), and later describe how the analysis
may be applied to arbitrary realistic gate geometries (Sec. \ref{sec:Arbitrary-gate-geometries}).
This interaction is described by \begin{equation}
\hat{V}_{E}=q\hat{E}\rho_{E}=q\hat{E}(x\cos\zeta+y\sin\zeta),\label{el-field}\end{equation}
 where $\hat{E}$ is the electric field amplitude, and $\rho_{E}$
denotes the projection of the electron's coordinate along the direction
$\zeta$ of the in-plane field. We have used a dipole-approximation,
assuming the electric field to be approximately constant across the
dot. This purely orbital interaction can lead to spin decay only when
combined with some spin-dependent part $\hat{V}_{\textrm{spin}}$
of the Hamiltonian (to be specified below). Thus, the total perturbation
added to $\hat{H}_{0}$ reads: \begin{equation}
\delta\hat{H}=\hat{V}_{\text{spin}}+\hat{V}_{E}.\label{perturbation}\end{equation}
Besides, the full Hamiltonian also contains a term describing the
electromagnetic environment, which determines the dynamics of the
electric field.

\section{Spin flip rate in second order Golden Rule}

\label{sec:Spin-flip-rate}According to Fermi's Golden Rule, the perturbation
$\hat{V}_{E}$ produces transitions at the rate \begin{eqnarray}
\Gamma=(2\pi/\hbar^{2})\left|\la f'|\,\rho_{E}\,|i'\ra\right|^{2}e^{2}\la\hat{E}\hat{E}\ra_{\omega}.\label{rate1}\end{eqnarray}
 Here $\hbar\omega=E_{i'}-E_{f'}$ is the energy absorbed by the environment,
and $\la\hat{E}\hat{E}\ra_{\omega}=(2\pi)^{-1}\int dt\, e^{i\omega t}\langle\hat{E}(t)\hat{E}\rangle$
is the spectrum of the electric field. The spectrum may be related
to the impedance of the gate circuit (Fig. \ref{fig:model}) by using
the Fluctuation-Dissipation Theorem (FDT) \cite{LL9}: \begin{eqnarray}
\la\hat{E}\hat{E}\ra_{\omega}=\pi^{-1}d^{-2}R(\omega)\hbar\omega\left(1-e^{-\hbar\omega/(k_{B}T)}\right)^{-1}.\label{nquistnoise}\end{eqnarray}
 Here $d$ is an effective distance determining the conversion between
voltage drop and electric field (distance between capacitor plates
in the simplest case; see Sec. \ref{sec:Arbitrary-gate-geometries}
for the general case). For the circuit of Fig. \ref{fig:model}, the
effective resistance responsible for the voltage fluctuations is given
by $R(\omega)=R/(1+(RC\omega)^{2})$. 

Spin-flip transitions are possible only because the matrix-element
in Eq. (\ref{rate1}) is to be evaluated in presence of the spin-dependent
part $\hat{V}_{\textrm{spin}}$, leading to a finite admixture of
different spin states. According to standard stationary perturbation
theory (leading to second order Golden Rule), we have: \begin{equation}
\la f'|\,\rho_{E}\,|i'\ra=\la f\!\!\down\!\!|\,\rho_{E}G(\E_{i\up})\hat{V}_{\text{spin}}+\hat{V}_{\text{spin}}G(\E_{f\!\down})\rho_{E}\,|i\!\!\up\ra.\label{amplitude}\end{equation}
 Here, the resolvent for $\hat{H}_{0}$ is $G(\E)=(\E-\hat{H}_{0})^{-1}$.
For the transition between the Zeeman sublevels of the orbital ground
state, we have $\E_{i\up}=\varepsilon_{0}+\hbar\omega/2=\E_{f\!\down}+\hbar\omega$,
where $\hbar\omega=|\text{g}\mu_{B}B|$ is the Zeeman energy, and
$\varepsilon_{0}$ is the energy of the orbital ground state. Eqs.
(\ref{rate1}) and (\ref{amplitude}) describe a second-order transition
from $|i\!\!\up\ra$ to $|f\!\!\down\ra$. Note that $\rho_{E}$ connects
the lateral orbital ground state $\phi_{00}$ to $\phi_{10}$ and
$\phi_{01}$ only.

We will now consider two specific spin-flip perturbations $\hat{V}_{\textrm{spin}}$.

\section{Hyperfine interaction}

\label{sec:Hyperfine-interaction}The hyperfine contact interaction
of an electron spin $\hat{\mathbf{S}}$ with nuclear spins $\hat{\mathbf{I}}_{j}$
at positions ${\mathbf{R}}_{j}$ has the form {[}\onlinecite{DPreview}{]}\begin{eqnarray}
\hat{V}_{\text{HF}}={\sum}_{j}v_{0}A_{j}\,\hat{{\mathbf{S}}}\cdot\hat{{\mathbf{I}}}_{j}\;\delta({\mathbf{r}}-{\mathbf{R}}_{j}),\label{HF-interaction}\end{eqnarray}
 where the $A_{j}$ are the hyperfine coupling constants (depending
on species), and $v_{0}$ is the volume of the unit cell. We will
neglect the nuclear Zeeman splitting.

In the limit of small Zeeman splitting, $\omega\ll\Omega_{0}$, we
obtain for the amplitude (\ref{amplitude}): \begin{align}
 & \la f'|\,\rho_{E}\,|i'\ra_{\text{HF}}=-2lv_{0}\frac{\left(1+\omega_{c}^{2}/(2\Omega_{0})^{2}\right)^{1/2}}{\hbar\Omega_{0}}{\sum}_{j}A_{j}\nn\label{amplitude2}\\
 & \times\la\down\!\!|\hat{S}^{\al}|\!\!\up\ra\,\la f_{N}|\hat{I}_{j}^{\al}|i_{N}\ra\Phi(z_{j},\rho_{j})\cos(\zeta-\varphi_{j}).\end{align}
 The initial and final states of the nuclear spin system are denoted
$i_{N},\, f_{N}$. We have defined $\Phi(z_{j},\rho_{j})=\chi_{0}^{2}(z_{j})\phi_{00}(\rho_{j})|\phi_{10}(\rho_{j})|$,
for the nuclei positions ($\rho_{j},\varphi_{j},z_{j}$), and $\chi_{0}(z)$
is the ground state of motion along the $z$-direction.

Inserting into Eq. (\ref{rate1}) and averaging over an ensemble of
unpolarized uncorrelated nuclear spins, we obtain the following expression
for the electron spin relaxation rate due to Nyquist noise and hyperfine
coupling: \begin{equation}
\Gamma_{\text{HF}}\!=\!\frac{4\pi}{3}I(I+1)\frac{A^{2}}{(\hbar\Omega_{0})^{2}}\frac{v_{0}\eta}{z_{0}d^{2}}\frac{R(\omega)\omega}{R_{Q}}\frac{\left(1+\omega_{c}^{2}/(2\Omega_{0})^{2}\right)}{1-e^{-\hbar\omega/(k_{B}T)}},\label{hf-rate}\end{equation}
 where $A^{2}={\sum}_{j}A_{j}^{2}$ with summation over all nuclei
in the unit cell, $R_{Q}=h/e^{2}$ is the quantum of resistance, $\eta=z_{0}l^{2}\!\int\!\Phi(z,\rho)^{2}dzd{\bm\rho}$
is a dimensionless form factor involving an average over nuclei positions
$(z,\rho,\varphi)$, and $z_{0}$ is the 2DEG thickness.

In the following, we provide numerical estimates using typical parameters
for GaAs QDs: hyperfine coupling constant\cite{Merkulov} $A^{2}\simeq1.2\times10^{-3}$
meV$^{2}$, nuclear spin value $I=3/2$, unit cell volume $v_{0}=(5.65\;\text{\AA})^{3}$,
effective electron mass $m=0.067m_{0}$, $|\text{g}\mu_{B}|=0.025$
meV T$^{-1}$, $\hbar\omega_{c}/B_{z}=1.76$ meV T$^{-1}$, $z_{0}=10$
nm, and a geometrical factor of $\eta=9/(16\cdot4\pi)$ for the approximate
solution of an inversion layer (triangular well) potential (see, e.g.,
Ref. \onlinecite{Ando82}).

Thus, the numerical value of the electron spin-flip rate is: \begin{align}
\Gamma_{\textrm{HF}} & \approx0.6\,\text{Hz}\times\frac{R(\omega)}{R_{Q}}\left(\frac{1\mu\text{m}}{d}\right)^{2}\left(\frac{1\text{meV}}{\hbar\Omega_{0}}\right)^{2}\frac{B}{1\text{T}}\nn\label{rate4}\\
 & \times\frac{1+0.8\times\left(\frac{1\,\textrm{meV}}{\hbar\Omega_{0}}\frac{B_{z}}{1\text{T}}\right)^{2}}{1-\exp\left[-0.3\times\frac{B}{1\text{T}}\frac{1\text{K}}{T}\right]}.\end{align}
 Concerning $R(\omega)=R/(1+(RC\omega)^{2})$, a lower bound for the
cutoff frequency $1/(RC)$ is determined by the charging energy $E_{c}$
of the dot (involving the total capacitance $C_{\textrm{tot}}>C$),
in the form $(E_{c}/\hbar)(R_{Q}/R)$. For current GaAs experiments,
a typical value of $E_{c}=1$ meV yields $RC\omega\sim10^{-1}(R/R_{Q})(B/1\text{T})$,
such that $R(\omega)$ deviates from $R$ only at rather high magnetic
fields, as long as $R<R_{Q}$. As a function of $R$, the maximum
relaxation rate is reached at $R=1/(C\omega)$. In typical GaAs quantum
dots this corresponds to $R_{\text{max}}=40R_{Q}(1\text{T}/B)$.

For a confinement frequency of $\hbar\Omega_{0}=1$ meV, the rate
is equal to about $0.1\,\textrm{Hz}$ at $T=1$ K and $B=0.1$ T,
with $R/R_{Q}=10^{-2}$, $d=0.5\,\mu\textrm{m}$. This is comparable
to electron spin relaxation due to hyperfine interaction involving
piezoelectric phonons \cite{Erlingsson02} at the same values of $\Omega_{0}$,
$T$ and $B$. Since the Nyquist mechanism considered here yields
a zero-temperature rate linear in transition frequency (instead of
cubic as in Ref. \onlinecite{Erlingsson02}), it dominates at smaller
magnetic fields $B<0.1\,\textrm{T}$. The relaxation rate can reach
large values if the resistance approaches $R_{\text{max}}$; e.g.
at $T=1$ K and $B=1$ T (with $R_{\text{max}}=40R_{Q}$), we would
have $\Gamma_{\text{HF}}\simeq0.3$ kHz, much larger than the rate
due to electron-phonon coupling \cite{Erlingsson02}.

If new electrons are supplied to the dot in a transport situation
or an additional electron spin relaxation mechanism is effective,
then the present mechanism may also result in nuclear spin relaxation,
with a rate equal to $\Gamma_{\textrm{HF}}/N$, with $N$ the number
of nuclei. For the parameters used here (with $N\simeq5\times10^{5}$),
and at $T=1$ K, $R=10^{-2}R_{Q}$, the nuclear relaxation rate turns
out to be negligibly small ($10^{-7}$ Hz). In comparison, a recent
transport experiment in GaAs quantum dots \cite{Huettel03} found
nuclear spin relaxation times of the order of 10 minutes at $T=100$
mK and $B=40$ mT \textbf{}(cf. Ref. \onlinecite{geller} for relevant
theory). In order for the present Nyquist mechanism to yield a comparable
rate, a resistance $R\sim R_{\textrm{max}}\sim10^{3}R_{Q}$ would
be required.

\section{Spin-orbit interaction}

\label{sec:Spin-orbit-interaction}The other, numerically much more
important spin-relaxation mechanism we will consider is due to the
combination of Nyquist noise and spin-orbit coupling: \begin{equation}
\hat{V}_{\text{SO}}=\alpha(\hat{\pi}_{x}\hat{\sigma}_{y}-\hat{\pi}_{y}\hat{\sigma}_{x})+\beta(\hat{\pi}_{y}\hat{\sigma}_{y}-\hat{\pi}_{x}\hat{\sigma}_{x}).\label{SO1}\end{equation}
 The first term in (\ref{SO1}) is the Rashba term arising from any
structural inversion asymmetry of the heterostructure, the second
is the (linear) Dresselhaus term, a consequence of the bulk inversion
asymmetry of the semiconductor material. The main crystallographic
axes are assumed to be aligned with $x,y,z$.

By means of the commutation relations $(i\hbar/m)\hat{{\bm\pi}}=[{\bm\rho},\hat{H}_{0}]$
and $[\hat{H}_{0},\hat{{\bm\sigma}}]=i\text{g}\mu_{B}\hat{{\bm\sigma}}\times{\mathbf{B}}$,
we get: \begin{equation}
\hat{V}_{\textrm{SO}}=-i[{\bm\xi}\hat{{\bm\sigma}},H_{0}]+\text{g}\mu_{B}\hat{{\bm\sigma}}[{\mathbf{B}}\times{\bm\xi}],\label{SO3}\end{equation}
 where we have introduced the vector ${\bm\xi}=(m/\hbar)(-\beta x-\alpha y,\alpha x+\beta y,0)$.

The contribution of the first term in Eq. (\ref{SO3}) to the amplitude,
Eq. (\ref{amplitude}), yields $[\rho_{E},{\bm\xi}\hat{{\bm\sigma}}]=0$.
Thus, we obtain, in the limit $\omega\ll\Omega_{0}$: \begin{equation}
\la f'|\,\hat{\rho}_{E}\,|i'\ra_{\text{SO}}=\frac{\sqrt{\alpha^{2}+\beta^{2}}}{\hbar\Omega_{0}^{2}}\,\text{g}\mu_{B}\la f\!\!\down\!\!|\hat{{\bm\sigma}}[{\mathbf{B}}\times{\mathbf{\Omega}}]|i\!\!\up\!\ra,\label{amplitude-so}\end{equation}
 Here, we have defined the vector ${\mathbf{\Omega}}=(\cos(\zeta-\gamma),-\sin(\zeta+\gamma),0)$,
where $\gamma$ depends on the ratio of spin-orbit coupling parameters:
$\tan\gamma\equiv\alpha/\beta$.

\begin{figure}
\includegraphics[%
  width=0.50\columnwidth]{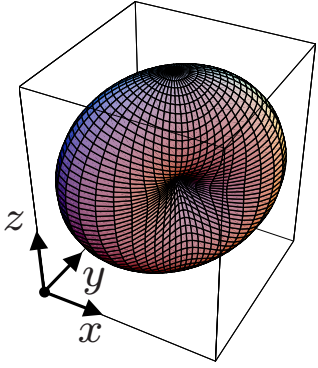}

\caption{(Color online) Angular dependence of the spin relaxation rate $\Gamma_{\textrm{SO}}$
on magnetic field direction $(\theta,\varphi)$, according to Eq.
(\ref{angular-dep}), for $\varphi^{*}=-\pi/4$. The shape will be
rotated around the z-axis for different values of $\varphi^{*}$,
which is determined by the electric field direction $\zeta$ and the
spin-orbit coupling constant ratio expressed via $\gamma$.}

\label{fig:angular-dependence}
\end{figure}

We parameterize the direction of the magnetic field as ${\mathbf{B}}=B(\sin\theta\cos\varphi,\sin\theta\sin\varphi,\cos\theta)$.
Then the electron spin relaxation rate due to spin-orbit coupling
combined with Nyquist noise is given by: \begin{align}
\Gamma_{\textrm{SO}}=4\pi\frac{\alpha^{2}+\beta^{2}}{d^{2}}\frac{R(\omega)}{R_{Q}}\,\frac{\omega^{3}}{\Omega_{0}^{4}}\,\frac{I(\zeta,\gamma\,;\theta,\varphi)}{1-e^{-\hbar\omega/(k_{B}T)}},\label{rate3}\end{align}
 where \begin{eqnarray}
I(\zeta,\gamma;\theta,\varphi) & = & (\sin^{2}(\zeta+\gamma)+\cos^{2}(\zeta-\gamma))\nonumber \\
 &  & \times\left(\cos^{2}\theta\,+\sin^{2}\theta\,\sin^{2}(\varphi-\varphi^{*})\right)\label{angular-dep}\end{eqnarray}
 yields the angular dependence of the relaxation rate (see Fig.~\ref{fig:angular-dependence}),
with the angle $\varphi^{*}$ defined by the condition $\tan\varphi^{*}=-\sin(\zeta+\gamma)/\cos(\zeta-\gamma)$.

We note that, for fixed direction of the fluctuating electric field,
$I$ vanishes for \emph{in-plane} magnetic fields pointing into the
direction $\varphi^{*}$, i.e. $I(\zeta,\gamma\,;\pi/2,\varphi^{*})=0,\;{\bm\forall}\,\zeta,\gamma$
(when ${\bm\Omega}\parallel\bm B$). In the particular case of equal
spin-orbit coupling constants, $\alpha=\pm\beta$ and thus $\gamma=\pm\pi/4$,
the first factor in $I$ reduces to $2\sin^{2}(\zeta\pm\pi/4)$. This
means the relaxation rate may even vanish regardless of magnetic field
direction, provided the electric field across the dot points along
$\zeta=\mp\pi/4$, thus $I(\mp\pi/4,\pm\pi/4\,;\theta,\varphi)=0,\;{\bm\forall}\,\theta,\varphi$
(since ${\bm\Omega}=0$ in Eq. (\ref{amplitude-so})). In the case
of an arbitrary gate geometry (treated in Sec. \ref{sec:Arbitrary-gate-geometries}),
the rate usually does not vanish any more but can still have some
pronounced directional dependence.

Conversely, for $\alpha=\pm\beta$ the rate may also vanish for arbitrary
orientation $\zeta$ of the electric field, if the magnetic field
lies in the plane and $\varphi=\mp\pi/4$ , i.e. $I(\zeta,\pm\pi/4\,;\pi/2,\mp\pi/4)=0,\;{\bm\forall}\,\zeta$.
As pointed out in Ref. \onlinecite{SchliemannPRL03}, for $\alpha=\pm\beta$
the electron spin component $\hat{\sigma}_{x}\mp\hat{\sigma}_{y}$
commutes with the spin-orbit interaction $\hat{V}_{\text{SO}}$, and
for $\bm B\parallel(1,\mp1,0)$ it commutes with $\hat{H}_{0}$ as
well. Thus, the suppression of spin relaxation in this case is exact
(see also Ref. \onlinecite{Golovach03}).

The numerical value for the relaxation rate is: \begin{align}
 & \Gamma_{\textrm{SO}}\approx0.4\,\text{kHz}\times\frac{R(\omega)}{R_{Q}}\left(\frac{1\mu\text{m}}{d}\right)^{2}\left(\frac{1\text{meV}}{\hbar\Omega_{0}}\right)^{4}\left(\frac{B}{1\text{T}}\right)^{3}\nn\label{rate-so-numerical}\\
 & \times\left(\frac{1\mu\textrm{m}}{\lambda_{\textrm{SO}}}\right)^{2}\, I(\zeta,\gamma\,;\theta,\varphi)\left[1-\exp\left[-0.3\times\frac{B}{1\text{T}}\frac{1\text{K}}{T}\right]\right]^{-1},\end{align}
 with $1/\lambda_{SO}^{2}\equiv(m/\hbar)^{2}(\alpha^{2}+\beta^{2})$.

The rate $\Gamma_{\textrm{SO}}$ is about $1$~Hz at $T=1$~K, for
$B=0.1$~T, using the same parameters as for hyperfine interaction
(and $\lambda_{\textrm{SO}}=1\mu\textrm{m}$, $I(\zeta,\gamma\,;\theta,\varphi)=1$).
At these parameters, this is comparable with the relaxation rate due
to the combined action of spin-orbit and piezoelectric electron-phonon
interactions \cite{KhN01,Golovach03}, which, however, vanishes like
$B^{5}$, instead of the $B^{3}$ dependence we have found here, such
that the Nyquist mechanism dominates for lower fields. While a $B^{3}$-dependence
was also found in Ref. \onlinecite{KhN01} for a two-phonon relaxation
process, the efficiency of that process decreases drastically when
the temperature becomes lower than about $1$~K. 

Very recently, the spin relaxation time in a one-electron quantum
dot has been measured directly, using a time-resolved single-spin
detection setup\cite{KouwenhovenNature04}. Previously, it had been
possible only to obtain lower bounds on this time (see Ref. \onlinecite{HansonPRL03}
for an example). At an external magnetic field of $8{\rm T}$, the
relaxation time was found to be on the order of $1{\rm msec}$. Using
parameters $\hbar\Omega_{0}=1$ meV, $d=0.5\,\mu\textrm{m}$, and
$\lambda_{\textrm{SO}}=1\,\mu\textrm{m}$, we indeed can obtain a
rate of about $10^{4}$ Hz by assuming a reasonable value of $R/R_{Q}=10^{-2}$
(lower values of $R$ are needed if the effective $d$ is smaller).
The observed magnetic field dependence is not inconsistent with a
$B^{3}$ contribution, although more data are needed in this regard.
Thus, the present Nyquist mechanism may be about as efficient as the
piezoelectric mechanism of Ref. \onlinecite{Golovach03}, for experimentally
relevant situations, and it would be interesting to see whether the
two effects can be distinguished in further measurements on similar
setups (see below). 

In contrast, we do not expect Nyquist noise to contribute to spin
relaxation rates in self-assembled quantum dots\cite{KrutvarNature},
which are apparently well explained by piezo-electric electron-phonon
coupling\cite{KhN01} alone.

As the rates $\Gamma_{\textrm{HF}}$ and $\Gamma_{\textrm{SO}}$ depend
on the relevant circuit resistance $R$, it might be interesting to
have a device where $R$ can be controlled (e.g. in a superconducting
gate circuit, with a strong decrease in $R(\omega)$ for $T<T_{c}$
and tunability by a magnetic flux; or using varactor diodes as tunable
capacitors in an $LC$-circuit). \textbf{}Inserting controlled dephasing
by a large $R$ could be a way to cross-check measurements of spin
relaxation rates (e.g. $R=4R_{Q}$ would yield an easily measurable
rate $\Gamma_{SO}$ on the order of $10^{7}$Hz, for the previous
parameters).

\begin{figure}
\includegraphics[%
  width=0.80\columnwidth]{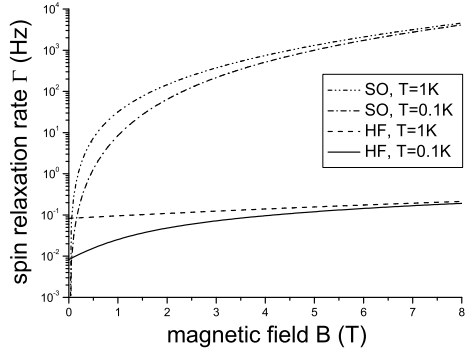}

\caption{Electron spin relaxation rate due to Nyquist noise, Eqs. (\ref{rate4})
and (\ref{rate-so-numerical}), at $\hbar\Omega_{0}=1$ meV, $R/R_{Q}=10^{-2}$,
$d=0.5\,\mu\textrm{m}$, $\lambda_{\textrm{SO}}=1\,\mu\textrm{m}$,
$\varphi=0$, $\theta=\pi/2$, $\zeta=0$, $\gamma=-\pi/4$. At low
fields, the hyperfine rate (HF) dominates over the spin orbit rate
(SO).}

\label{fig:rate-mfield}
\end{figure}

\section{Arbitrary gate geometries}

\label{sec:Arbitrary-gate-geometries}We now explain how our results
may be extended to arbitrary experimental configurations for gated
lateral dot structures (see Fig. \ref{cap:Typical-gated-lateral}).

In general, the dot experiences a fluctuating electric field which
is due to the fluctuating voltages $\hat{U}_{j}$ on the gates $j=1\ldots N_{G}$.
(We assume these voltages to be measured relative to ground) The relation
between the in-plane electric field components ($l=1,2$) at the center
of the dot and the voltages is linear, 

\begin{equation}
\hat{E}_{l}=\sum_{j=1}^{N_{G}}M_{lj}\hat{U}_{j}\,.\label{EMU}\end{equation}
The real-valued matrix $M$ depends on the geometry of the gates and
the dielectric substrate (it has to be obtained from an electrostatic
simulation for specific geometries). This matrix has the dimension
of an inverse length, and roughly corresponds to the factor $d^{-1}$
we introduced in Eq. (\ref{nquistnoise}) for our simple model system
(which can be described by $N_{G}=2$ and $M_{l2}=-M_{l1}$ in this
scheme). 

The gates are nodes in an electronic circuit that is described by
an impedance matrix $Z_{ij}$. Thus, according to the FDT, we have:

\begin{equation}
\left\langle \hat{U}_{j'}\hat{U}_{j}\right\rangle _{\omega}=\frac{1}{2\pi}\left[Z_{j'j}(\omega)+Z_{jj'}^{*}(\omega)\right]\frac{\hbar\omega}{1-e^{-\hbar\omega/T}}\,.\label{UUcorr}\end{equation}
We now make the reasonable assumption that the magnetic field is not
strong enough to appreciably influence the impedance matrix of the
circuit. Then the Onsager-Casimir relation $Z_{j'j}(B,\omega)=Z_{jj'}(-B,\omega)$,
evaluated at $B=0$, implies reciprocity $Z_{j'j}=Z_{jj'}$ and thus
$(Z_{j'j}(\omega)+Z_{jj'}^{*}(\omega))/2={\rm Re}Z_{j'j}(\omega)$.
Inserting Eq. (\ref{UUcorr}) into Eq. (\ref{EMU}), we have 

\begin{equation}
\left\langle \hat{E}_{l'}\hat{E}_{l}\right\rangle _{\omega}=\frac{\hbar\omega/\pi}{1-e^{-\hbar\omega/T}}\sum_{j'j}M_{l'j'}M_{lj}{\rm Re}Z_{j'j}(\omega)\,.\end{equation}
Reciprocity makes the matrix $\left\langle \hat{E}_{l'}\hat{E}_{l}\right\rangle _{\omega}$
real-valued and symmetric. Physically, this means there is no circular
component to the thermal electric field fluctuations. Thus, at given
$\omega$, we can diagonalize this matrix by choosing two orthogonal
directions in the 2DEG plane. The total spin relaxation rate becomes
equal to the sum of the rates for these two directions, obtained according
to our previous description. The angle $\zeta$ in our notation denotes
the direction of the given principal axis, and $\left\langle \hat{E}\hat{E}\right\rangle _{\omega}$
is equal to the corresponding eigenvalue. As a consequence, the total
spin-orbit rate $\Gamma_{SO}$ will not vanish completely for any
magnetic field direction, but an anisotropy will generally remain
due to the anisotropy of the gate geometry (or the impedance matrix).

However, even if the geometry were perfectly known, and the matrix
$M$ were calculated using a numerical solver of the Poisson equation,
it is currently difficult to go beyond estimates for realistic setups
(like, e. g., those used in the experiments by the Munich\cite{Huettel03}
or Delft \cite{HansonPRL03,KouwenhovenNature04} groups). This is
because the impedance matrix of the gate circuit (and thus the noise
properties) have never been studied in detail at the high frequencies
under consideration, and the nominal circuit diagrams are probably
valid only at low frequencies. Nevertheless, our estimate for the
effective resistance entering the Nyquist noise, $R/R_{Q}\sim10^{-2}$,
seems to be roughly comparable to the actual numbers (Munich group\cite{huettelprivat})
or at least cannot be ruled out at present (Delft group\cite{hansonprivat}).
In this context it should also be noted\cite{huettelprivat} that
part of the circuit is at higher temperatures, possibly contributing
more noise than estimated using the base temperature of the setup.

\section{Other dot shapes and beyond dipole approximation}

\label{sec:Other-dot-shapes}If we consider an anisotropic or, in
general, nonparabolic lateral confinement potential, the orbital single-particle
wave functions and energies will change. Most importantly, this results
in nonvanishing transition matrix elements between the ground state
and more than just two excited orbital states. The same is true if
we go beyond the dipole approximation, i.e. we no longer assume the
fluctuating electric field to be constant across the dot. 

Thus, the sum over intermediate states implicit in Eq. (\ref{amplitude})
extends over more states. However, generally speaking the qualitative
picture of our previous analysis does not change, and neither do the
quantiative estimates, provided the deviations from the parabolic
shape are not too large. This is because transitions via higher intermediate
states are suppressed anyway by larger energy denominators, and, in
particular, no extra channel via low-lying energies opens up.

\section{Conclusions}

In summary, we analyzed electron spin relaxation due to the combination
of Nyquist noise with both hyperfine and spin-orbit interaction. For
the case of spin-orbit coupling, a remarkable dependence on the directions
of magnetic and electric fields is observed. For equal spin-orbit
coupling constants (realizable by tuning the Rashba term\cite{SchliemannPRL03}),
the rate may vanish exactly for a particular in-plane magnetic field
direction (independent of gate geometry) or, for arbitrary angles
of the magnetic field, become very small provided the fluctuating
electric field points predominantly into a certain fixed direction.
The contribution to the spin relaxation rate may be as large as that
of the most important other mechanism (piezoelectric electron-phonon
coupling), for relevant experimental parameters. Furthermore, if a
larger (preferably tunable) gate resistance is relevant for a given
experimental setup, or microwave noise is deliberately applied to
the gate, the mechanism analyzed here may dominate over an extended
range of magnetic field values and might be more easily distinguished
from other mechanisms.

\begin{acknowledgments}
We thank R. Hanson, L. W. van Beveren, and A. Hüttel for sharing information
on their experimental setups and providing interesting feedback. One
of us, V. A. A., would like to thank A. V. Chaplik for useful discussions
of the results. F.M. has been supported by a DFG grant (MA 2611/1-1).
\end{acknowledgments}

\end{document}